\documentclass[10pt,amssymb,amsmath]{iopart}
\usepackage{graphicx,epsfig}
\begin{document}
\title{Semihard scattering  unraveled from collective dynamics at $\sqrt{s}$~=~17~GeV} 
%by azimuthal correlations}

\author{Jana Bielcikova\footnote[1]{jana.bielcikova@yale.edu} for the CERES/NA45 Collaboration
}

\address{ Physics Department, Yale University,  P.O. Box 208120, New Haven, CT 06520-8120, USA}

\begin{abstract}
We present a study of elliptic flow ($v_2$) and two-particle azimuthal correlations of charged hadrons 
and high transverse momentum ($p_T$) pions measured by the CERES experiment in Pb+Au collisions at $\sqrt{s}$~=~17~GeV. 
Azimuthal anisotropy $v_2$ increases linearly with $p_T$ to a value of 10$\%$ at $p_T$~=~1.5~GeV/$c$. Beyond this $p_T$ 
the slope decreases indicating a possible saturation at high $p_T$. Two-pion azimuthal 
anisotropies ($p_T>$~1.2~GeV/$c$) exceed $v_2$ by about 60$\%$ in semicentral collisions and
reveal non-harmonic contributions at close ($\Delta\phi\approx0$) and
back-to-back ($\Delta\phi\approx\pi$) angles that can be attributed to semihard
processes. While the close-angle peak remains unchanged,
the back-to-back peak is broadened and disappears in
central collisions. 
%Both components show only a weak preference towards the reaction plane.
%The results are discussed with the perspective of $p_T$
%broadening and jet quenching and related information from inclusive particle
%spectra at the SPS and RHIC.

\end{abstract}

%\pacs{25.75.Ld}


% \maketitle

\section{Introduction}
Elliptic flow is an important signature of collective dynamics in non-central heavy-ion collisions at
high energies. It is driven by anisotropic pressure gradients built up during the
early stage of the collision due to the geometrically anisotropic overlap zone 
of the colliding nuclei. Moreover, it carries information on such important issues 
as the equation of state (EOS) and the level of equilibration achieved. 
Elliptic flow manifests itself in an azimuthal anisotropy 
of particle yields with respect to the reaction plane.
Two-particle azimuthal correlations are sensitive to elliptic flow but at large transverse momenta ($p_T$) are expected
to reveal also relics of semihard scattering. 
%Therefore, a comparison of measurements of azimuthal anisotropies using the correlations with reaction plane and two-particle correlations might help to disentangle semihard scattering from collective flow. 
%The azimuthal anisotropy is a consequence of early partonic rescattering in
%the dense overlap zone of the nuclei at the time of collision and carries
%information on such important issues as the equation of state and the level
%of equilibration achieved. 
We report about successful attempts to trace primeval partonic scattering 
%and the onset of thermalization in the medium 
at  $\sqrt{s}$~=~17~GeV 
%together with and separated from 
by two-particle azimuthal correlations of pions at moderately large $p_T$ ($p_T>$1.2~GeV/$c$).  
%as a function of the centrality of the collision.

\section{Experimental setup and particle tracking}
\begin{figure}[t!]
\begin{center}
\includegraphics[width=10.5cm]{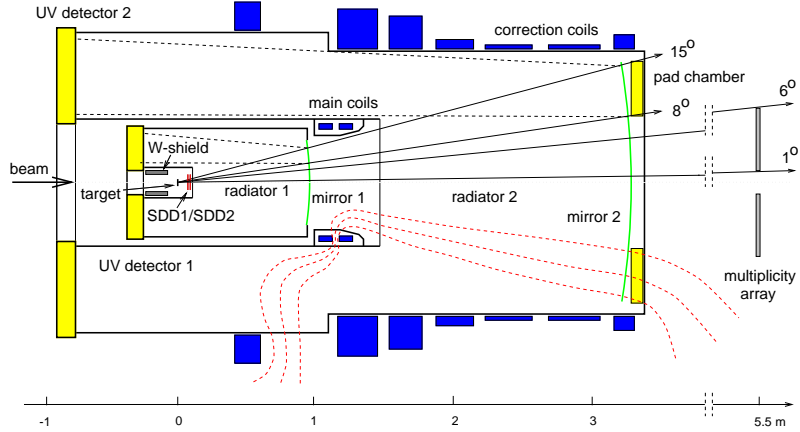}
\caption{CERES experimental setup in 1996.}
\label{expsetup} 
\end{center}
\end{figure}

\begin{figure}[h!]
\begin{tabular}{lr}
\includegraphics[height=6.0cm]{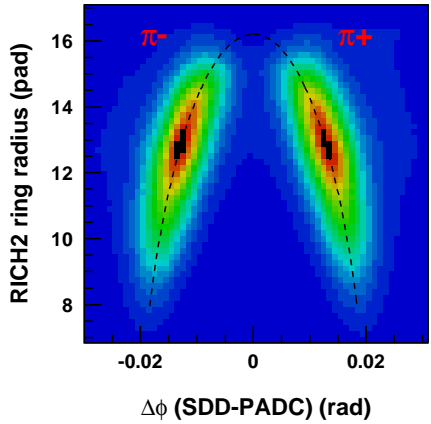}
&
\hspace{-2.5cm}
\begin{minipage}[t]{8.0cm}
\vspace{-5.5cm}
\caption{Correlation between the pion ring radius in RICH2 and the azimuthal deflection 
in the magnetic field measured between SDDs and PADC. The dashed line represents the expected correlation.}
\label{banana} 
\end{minipage}
\end{tabular}
\end{figure}

Figure~\ref{expsetup} shows the CERES experimental setup from 
1996. %before an upgrade by a time projection chamber. 
The spectrometer covers a pseudo-rapidity range 2.1~$<\eta<$~2.65 and has full 
azimuthal acceptance, which is important for studies of azimuthal 
distributions. Although the overall design was optimized to 
detect low-mass dilepton pairs~\cite{Lenkeit:1999xu}, CERES offers many capabilities 
in hadronic physics as well.   
Charged particle tracks are reconstructed on a statistical basis combining information 
from two radial silicon drift detectors (SDD1, SDD2) placed closely behind the target 
%with a hit measured by 
and a multi-wire proportional chamber (PADC) behind a magnetic field
used for momentum determination. %The currents in two superconducting solenoidal coils 
%producing the magnetic field have an opposite sign and result in a sharply located 
%azimuthal kick directly proportional to momentum.
%leaving the polar angle of a particle track to the first order unchanged. Momenta of charged particles are 
%determined from the deflection $\Delta\phi=\phi_{SDD}-\phi_{PADC}$ in the magnetic field:
%\begin{equation}
%\label{def1}    
%  p = \frac{96 \;{mrad}}{\Delta\phi} [{GeV}/c].
%\end{equation}
%used for their momentum determination. The currents in two superconduction solenoidal 
%coils which produce the magnetic field have an opposite sign and result in a sharply 
%located azimuthal kick leaving the polar angle of a particle track to the first order
%unchanged. The relative momentum resolution is influenced by the intrinsic detector resolution
%and at low momenta also by multiple scattering. It reaches value of ...
%is used for their momentum determination. A pair of superconducting solenoidal coils located between the
%two RICH detectors produces a magnetic field for the 
%momentum determination.
%The currents with an opposite sign in the two superconducting coils 
Charged pions are identified and distinguished from electrons by smaller ring radii
in two ring-imaging Cherenkov detectors (RICH1, RICH2). Since the RICH detectors are filled 
with CH$_4$ with a high Cherenkov threshold ($\gamma_{th}\simeq$~32), only pions with $p>$~4.5~GeV/$c$
produce Cherenkov light. % giving thus a natural selection of high-$p_T$ pions. 
Figure~\ref{banana} shows the correlation between the ring radius in RICH2 and the
azimuthal deflection in  magnetic field with the two islands corresponding to
negatively and positively charged pions, respectively. 
%The relative momentum resolution (Fig.~\ref{momresol}) is influenced by the intrinsic detector resolution
%and at low momenta also by multiple scattering. 
Pion momenta are determined from the ring radius measurement due to its higher precision in comparison to the deflection in magnetic field~\cite{Slivova:thesis}.
%as can be seen 
%from a comparison of both momentum determination methods in Fig.~\ref{momresol}.
%\begin{figure}[h!]
%    \begin{tabular}{lr}
%    \includegraphics[height=7.0cm]{momentum-rad-defl-paper.eps}
%    &
%    \hspace{-2.0cm}
%    \begin{minipage}[t]{8cm}   
%       \vspace{-6.5cm}
%       \caption{Momentum resolution of the CERES spectrometer determined from the deflection 
%        in magnetic field and from the ring radius measurement in the RICH detectors.}
%       \label{momresol}
%    \end{minipage}
%  \end{tabular}
%\end{figure} 

%The particle tracking efficiency $\epsilon$ is obtained from embedding of tracks into real events. We have found that
%$\epsilon$(h$^\pm$) reaches on average 55\% and decreases only weakly with increasing 
%centrality reflecting a very good position resolution of SDD and PADC detectors. However, pion reconstruction 
%efficiency is more complicated because both RICH detectors come into play. It depends of $N_ch$, $\theta$, 
%and momentum $p$. In the vicinity of the Cherenkov threshold it rises steeply with increasing momentum 
%followed by a plateau for $p=10-15$~GeV/$c$. 

\section{Centrality determination}

\begin{figure}[t!]
\begin{tabular}{lr}
\hspace{-0.8cm}
\includegraphics[height=6.4cm]{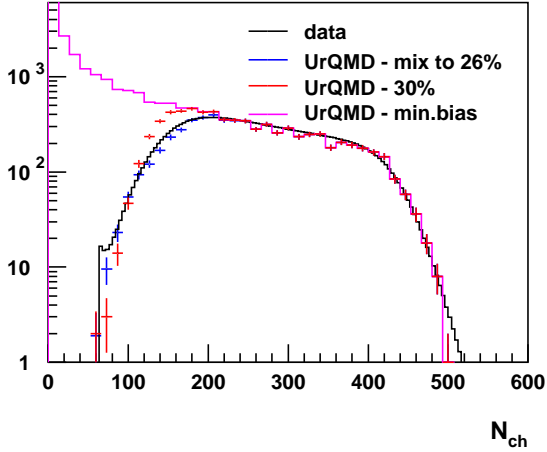}
&
\hspace{-2.8cm}
\begin{minipage}[t]{7.0cm}
\vspace{-6.0cm}
\caption{Charged particle multiplicity $N_{ch}$ measured by the silicon drift detectors  in the pseudo-rapidity interval
2~$<\eta<~$3. The data are corrected for efficiency losses and compared to UrQMD calculations for various 
centrality selections as given in the legend. An overall multiplicative factor of 1.03 was applied to $N_{ch}$ values 
from UrQMD in order to describe the data in central collisions.} 
%The data sample is divided into 6 centrality classes C1,C2,...,C6 schematically depicted 
%by horizontal lines. $N_{ch}$ is corrected for efficiency losses due to dead regions and pile-up.}
\label{multiplicity}
\end{minipage}
\end{tabular}
\end{figure}

We have analyzed $41\cdot10^6$ Pb+Au collisions taken at $\sqrt{s}$~=~17~GeV.
%at the most central (26$\pm$1.5)$\%$ of the geometric cross section $\sigma_{geo}$.
The centrality was determined offline using the number of charged particles $N_{ch}$ 
measured by the SDD in 2~$<\eta<$~3. The $N_{ch}$ distribution corrected for efficiency losses 
%due to dead regions, pile-up, and ghost tracks 
is shown in Fig.~\ref{multiplicity}. As the multiplicity detector 
used as a centrality trigger suffered from voltage 
instabilities, which were unfortunately not continously monitored, 
we had to use the UrQMD model~\cite{urqmd} to estimate the fraction of geometrical cross section $\sigma_{geo}$ 
measured. We have concluded that the measured data sample corresponds to the most central 
(26.0$\pm$1.5)$\%$ of $\sigma_{geo}$. %$rather than 30$\$ quoted earlier~\cite{Agakishiev:1998wu-Lenkeit:1999xu}. %The error of 1.5$\%$ reflects the fact that various centrality mixtures describe the $N_{ch}$ distribution.
The data sample was divided into 6 centrality classes summarized in Table~\ref{centrality-selection} together with the corresponding fraction of geometrical cross section $\sigma/\sigma_{geo}$, impact parameter $b$, number of participants $N_{part}$, and binary collisions $N_{coll}$
obtained from a Glauber calculation neglecting fluctuations~\cite{Eskola:1989yh}.
\begin{table}[h!]
\begin{center}
\begin{tabular}{|c|c|c|c|c|c|c|} \hline
Class & Events (10$^6$)& $N_{ch}$ & $\sigma/\sigma_{geo} (\%)$ & $b$ (fm) & $N_{part}$ & $N_{coll}$ \\ \hline \cline{1-7}  
%Class & Events (10$^6$) & $\langle N_{\text{ch}}\rangle$ & $\sigma/\sigma_{\text{geo}} (\%)$ 
%& $b$(fm) & $\langle N_{\text{part}}\rangle$ & $\langle N_{\text{coll}}\rangle$ \\ \hline  \cline{1-7}    
1&  7.77 & 147 &   21 - 26     &  6.8 - 7.5  & 159  & 293       \\ \hline
2&  6.58 & 198 &   17 - 21     &  6.0 - 6.8  & 189  & 368       \\ \hline
3&  5.66 & 234 &   13 - 17     &  5.3 - 6.0  & 222  & 453       \\ \hline
4&  6.06 & 273 &   9 - 13      &  4.4 - 5.3  & 255  & 542       \\ \hline
5&  6.05 & 321 &    5 - 9      &  3.4 - 4.4  & 289  & 639       \\ \hline
6&  8.16 & 395 &    $<$ 5      &  $<$ 3.4    & 336  & 774       \\ \hline
\end{tabular}
\caption{Definition of centrality classes.}
\label{centrality-selection}
\end{center}
\end{table}

\section{Collective elliptic flow}
The strength of elliptic flow is commonly quantified by the second Fourier coefficient $v_2$~\cite{Poskanzer:1998} of azimuthal particle distributions with respect to the reaction plane $\Psi_R$
\begin{equation}
\frac{dN}{d(\phi-\Psi_R)}=A(1+\sum_{n=1}^\infty 2\; v_n \cos (n(\phi-\Psi_R))).
\label{single-rp}
\end{equation}
{\it A priori}, the reaction plane is unknown and is therefore estimated on an event-by-event basis from charged particle tracks measured by 
the SDDs using a subevent method. Non-uniformities in the event plane distribution are removed by standard procedures~\cite{Poskanzer:1998}. Depending on centrality of a given collision, the r.m.s. of the event plane resolution is 35-40 degrees. 

\begin{figure}[t!]
\begin{tabular}{lr}
\hspace*{-0.4cm}
\includegraphics[height=5.5cm]{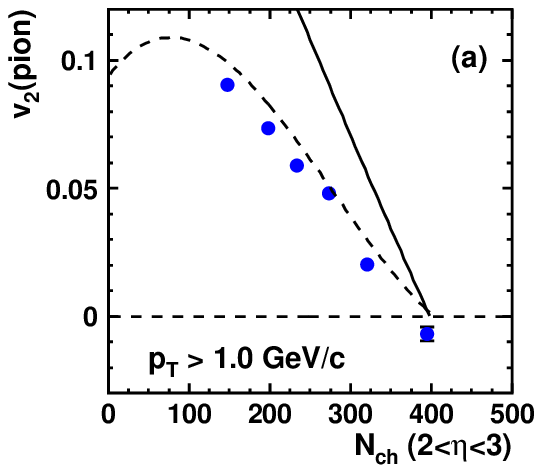}
&
\hspace*{-0.8cm}
\includegraphics[height=5.5cm]{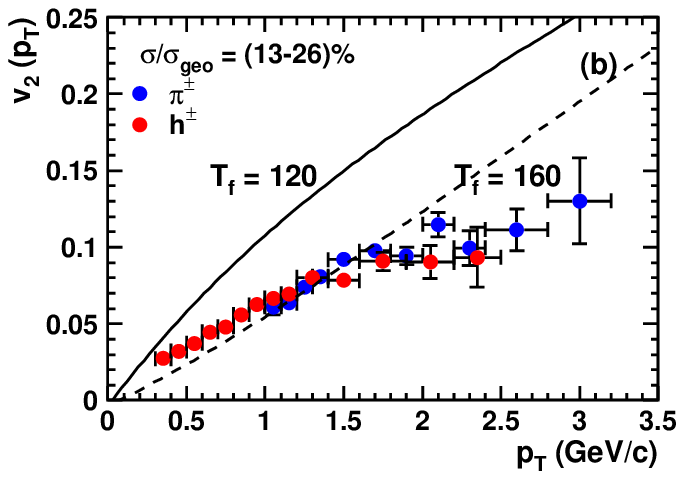}
\end{tabular}
\caption{Centrality (a) and $p_T$ dependence (b) of $v_2$ for charged pions and hadrons as indicated in the legend. Hydrodynamical calculations~\cite{Huovinen} with a phase transition at $T_c$~=~165~MeV are shown for kinetic freeze-out temperature $T_f$~=~120~MeV (solid line) and $T_f$~=~160~MeV (dashed line). 
The quoted errors are statistical only. The absolute systematic errors vary between 0.5$\%$ and 1.5$\%$ going from semicentral to central collisions.}
\label{v2-nch-pt}
\end{figure}

The centrality and $p_T$ dependence of $v_2$ corrected for the event plane dispersion is shown in Fig.~\ref{v2-nch-pt}. In addition, the $v_2(p_T)$ data points are corrected for Bose-Einstein correlations~\cite{Slivova:thesis,Dinh:1999mn,Adamova:2002wi}. Since this correction procedure becomes questionable for central collisions, the centrality dependence of $v_2$ was left uncorrected. 
We observe that $v_2$ decreases approximately linearly with centrality and vanishes in the most central collisions with no remaining asymmetry in the overlap zone present (Fig.~\ref{v2-nch-pt}a). The $p_T$ dependence of $v_2$ in semicentral collisions (Fig.~\ref{v2-nch-pt}b) shows a linear rise below $p_T$~=~1.5~GeV/$c$. Beyond $p_T\approx$~1.5~GeV/$c$ the slope decreases, possibly indicating a saturation of $v_2$ at high $p_T$ similar to observations at RHIC~\cite{RHIC-combined}.

A direct quantitative comparison with a hydrodynamical calculation~\cite{Huovinen} using an EOS with a first order transition to a quark gluon plasma at temperature $T_c$~=~165~MeV favors a higher freeze-out temperature $T_f$~=~160~MeV rather than a lower one, $T_f$~=~120~MeV. However, $T_f$~=~120~MeV is necessary in order to describe the $p_T$ spectra of protons. Possible explanations might be either  incomplete thermalization or a necessity to include viscous effects into calculations~\cite{Teaney}.

\section{Azimuthal correlations at high-p$_T$}
\begin{figure}[t!]
\begin{center}
\includegraphics[height=5.8cm]{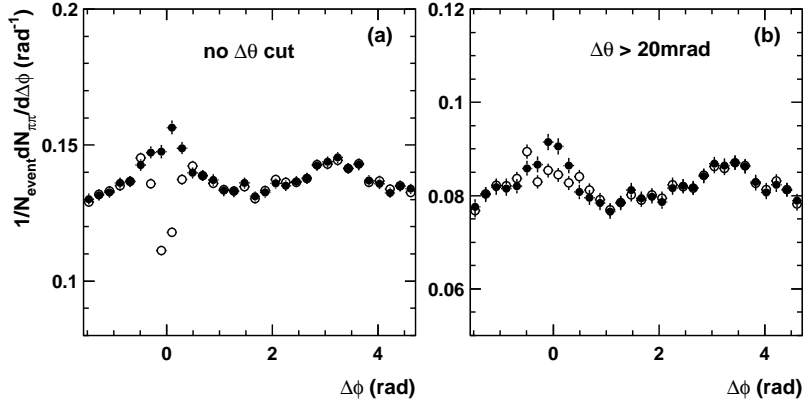}
\caption{Two-pion azimuthal distributions ($p_{{T}}>$1.2~GeV/$c$)
for the centrality class C1 (a) without and  (b) with a $\Delta\theta>$~20~mrad cut 
applied. Open symbols: data only after the MC correction for single track reconstruction 
efficiency. Closed symbols: data after an additional MC correction of the finite 
two-track resolution.}
\label{dtheta-0-20mrad-corrected}
\end{center}
\end{figure}

We turn to the measurement of two-particle azimuthal correlations of high-$p_T$ pions. The two-particle distributions in semicentral collisions corrected for single-track reconstruction efficiency are shown in Fig.~\ref{dtheta-0-20mrad-corrected}. At small opening angles, overlapping rings in the RICH detectors cause a drop in pair reconstruction efficiency which manifests itself as a dip around $\Delta\phi\approx$~0 (Fig.~\ref{dtheta-0-20mrad-corrected}a). This instrumental effect can be cured by using either a Monte-Carlo (MC) correction or by enforcing a full ring separation by imposing a cut in polar angle difference $\Delta\theta$. We have made a compromise between the two methods and use the separation cut $\Delta\theta>$~20~mrad (Fig.~\ref{dtheta-0-20mrad-corrected}b) which still keeps about 60$\%$ of pion data sample while reducing sensitivity to the MC correction by a factor of four. The corrected distributions reveal a strong anisotropy with maxima at close ($\Delta\phi\approx$~0) and back-to-back ($\Delta\phi\approx\pi$) angles. %The correction procedure is supported by the fact that anisotropies remain essentially unchanged whether or not the $\Delta\theta$ cut is applied.

\begin{figure}[b!]
\begin{center}
\begin{tabular}{lr}
\hspace*{-0.3cm}
\includegraphics[height=6.0cm]{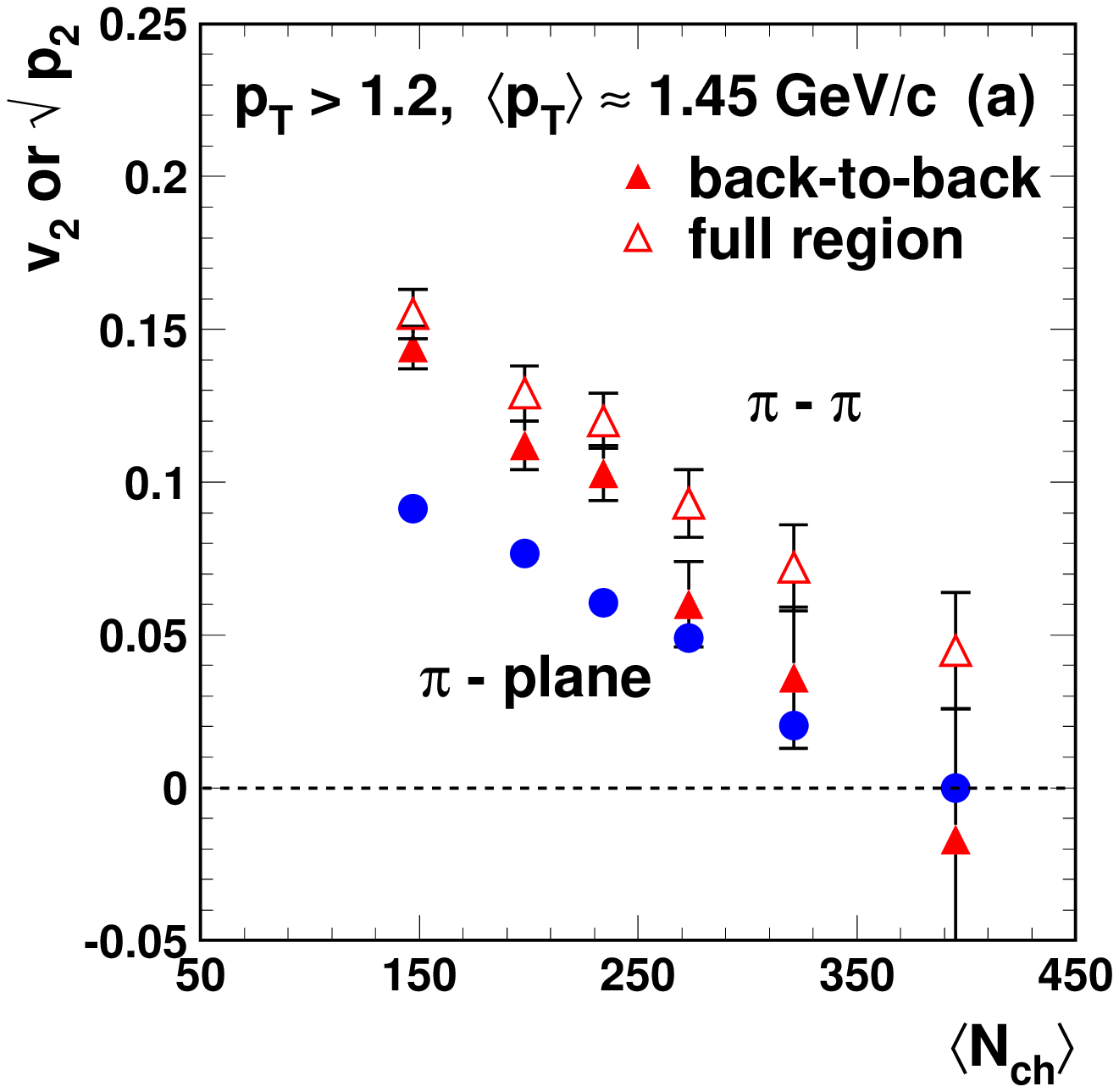}
&
\hspace*{-0.7cm}
\includegraphics[height=6.0cm]{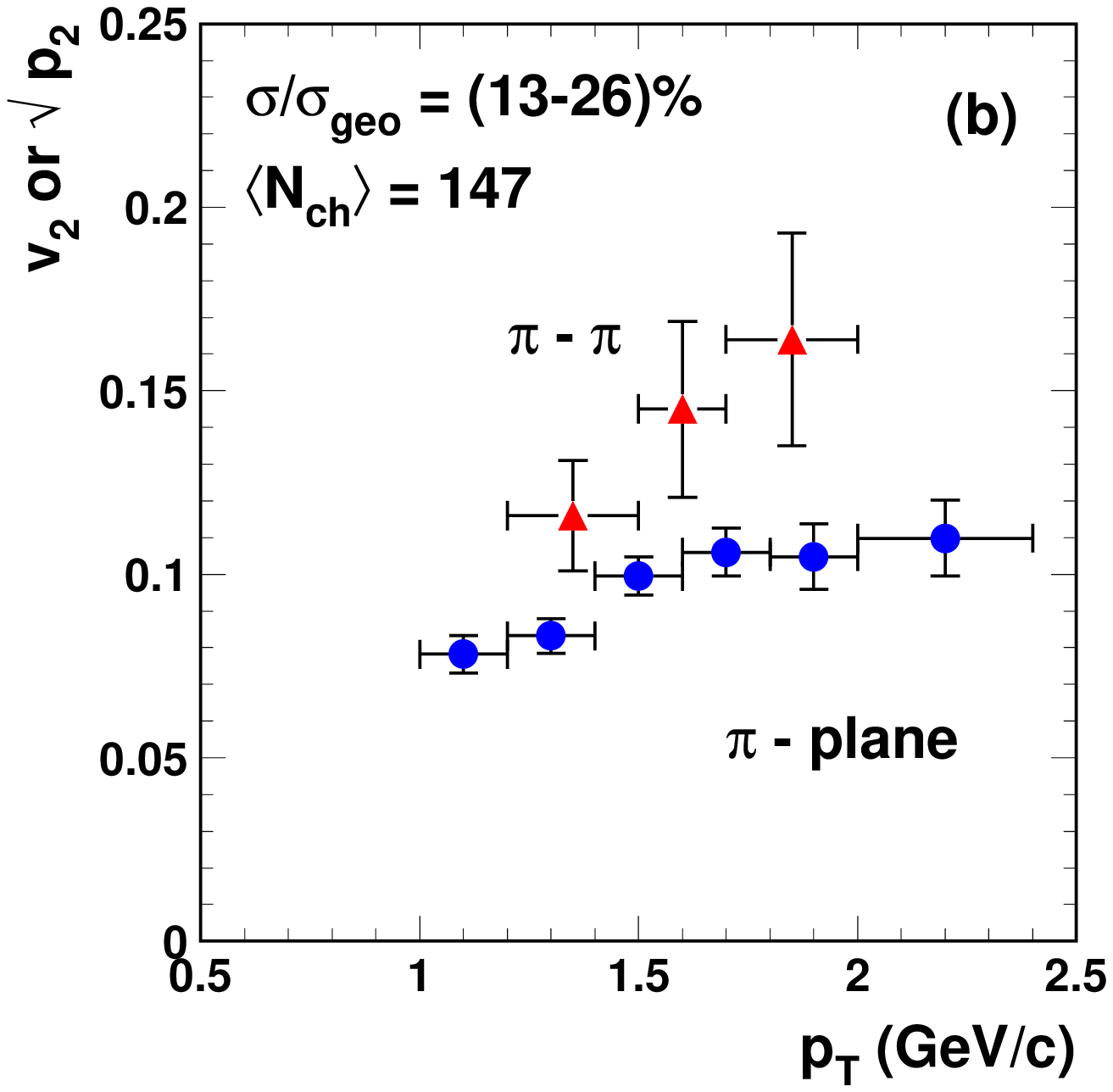}
\end{tabular}
\caption{Comparison of $v_2$ (circles) obtained from the event plane method and $\sqrt{p_2}$ (triangles) 
from two-pion azimuthal correlations. (a) Centrality dependence for full azimuth (open triangles) and a range
restricted to the back-to-back peak ($|\Delta\phi|\ge$~0.6~rad, closed triangles). (b) $p_T$ dependence for the
centrality class C1.}
\label{v2-p2-comparison}
\end{center}
\end{figure}
The two-particle distributions can be decomposed again using a Fourier method
\begin{equation}
\frac{dN}{d\Delta\phi} = B(1+\sum_{n=1}^\infty 2\; p_n \cos (n\Delta\phi)),
\label{2part-eq}
\end{equation} 
where $\Delta\phi$ is an azimuthal angle difference between any two pions in a given event. If only correlations due to collective flow are present, then $p_n=v_n^2$.
Figure~\ref{v2-p2-comparison} shows a comparison of the $\sqrt{p_2}$ values with $v_2$ obtained from the event plane method. We observe that the $\sqrt{p_2}$ values are systematically larger than $v_2$. A closer look at the centrality dependence shows that the anisotropy in the back-to-back region approaches zero for central collisions while the close-angle correlations persist even in the most central collisions. The gap between $v_2$ and $\sqrt{p_2}$ seems to increase with increasing $p_T$ (Fig.~\ref{v2-p2-comparison}b). Unfortunately, the statistical significance of this measurement is degraded by invoking a two-dimensional window in $p_T$.

\begin{figure}[h!]
\begin{center}
\begin{tabular}{lr}
\hspace*{-0.3cm}
\includegraphics[height=6.0cm]{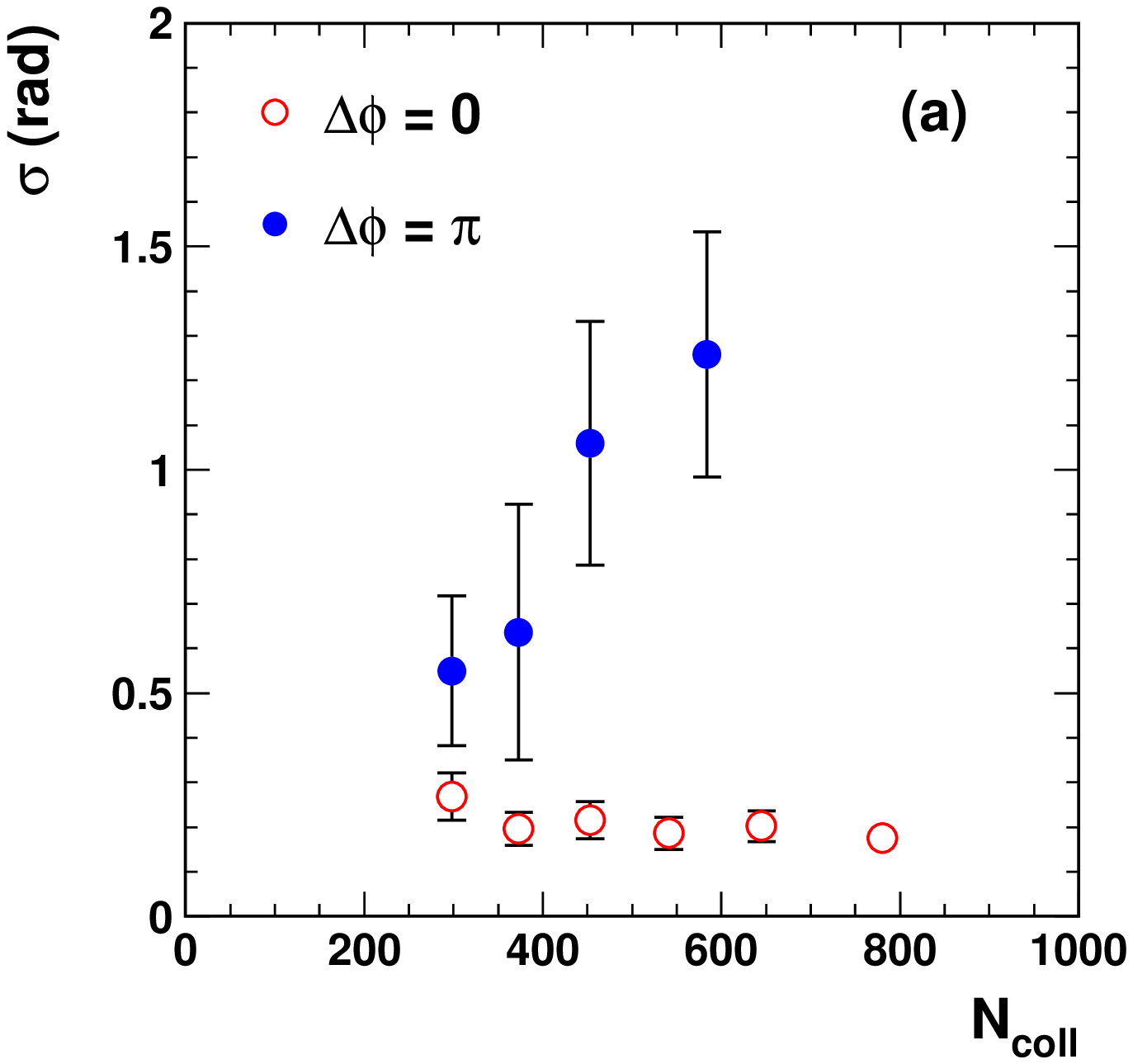}
&
\hspace*{-0.7cm}
\includegraphics[height=6.0cm]{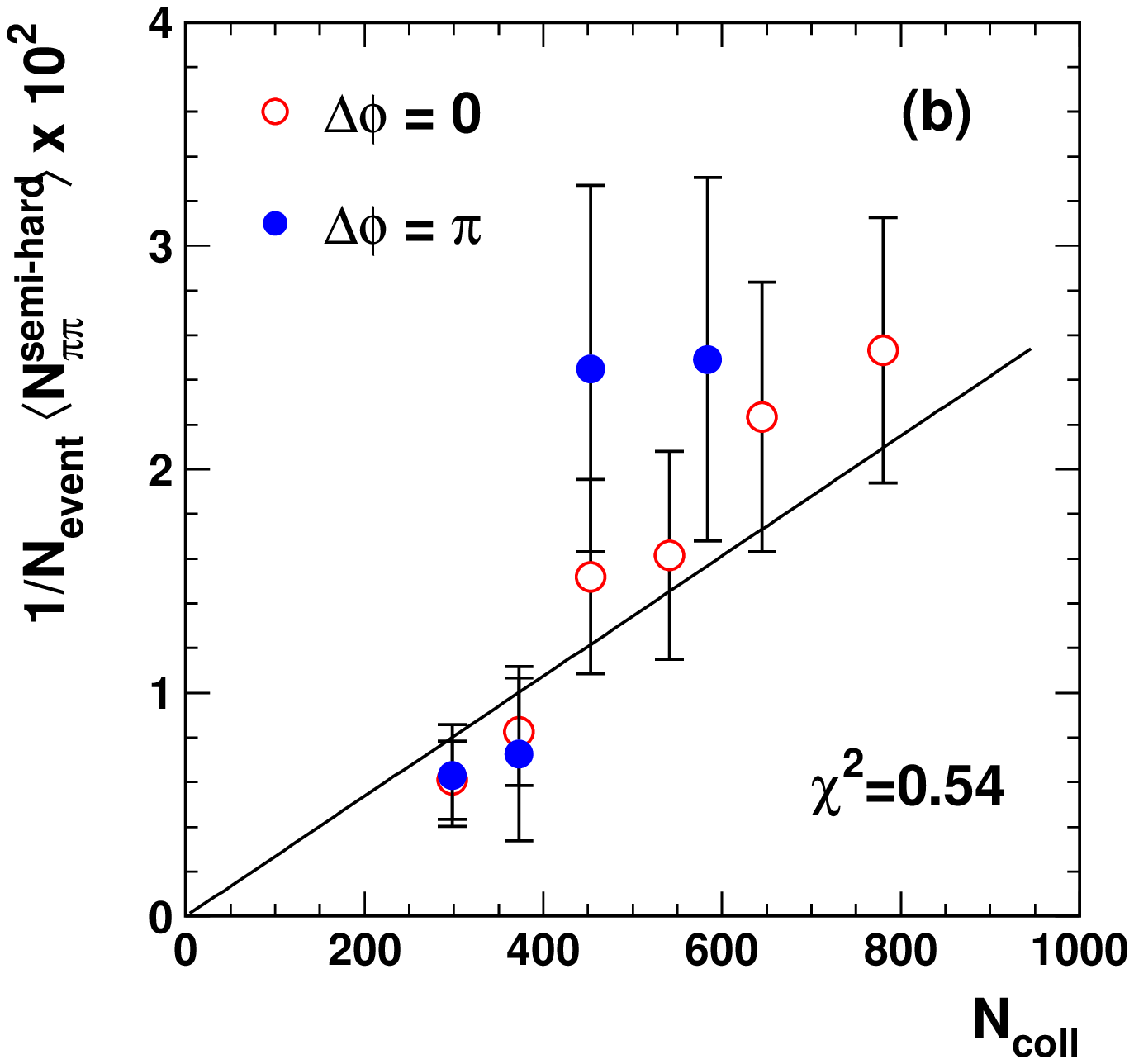}
\end{tabular}
\caption{Centrality dependence of (a) the Gaussian widths and (b) the yield  of close-angle (open symbols) and back-to-back (full symbols) correlation peaks for pions with $p_T>$~1.2~GeV/$c$.}
\label{sigma-yield}
\end{center}
\end{figure}
Assuming the observed excess is due to correlations of semihard origin, we have fit the distributions with two Gaussian peaks at $\Delta\phi=0$ and $\pi$ on top of the elliptic flow modulated background. The fit parameters are the Gaussian amplitudes, widths and background, while $v_2$ is fixed from the event plane method. The close-angle peak remains narrow (Fig.~\ref{sigma-yield}a) at $\sigma_0$~=~(0.23$\pm$0.03)~rad averaged over the measured centrality. The corresponding average momentum perpendicular to the partonic transverse momentum~\cite{Rak:2004gk} is 
%$\langle j_{T} \rangle = \pi/2\langle|j_{Ty}|\rangle\approx\sqrt{\pi}/2\langle p_T\rangle\sigma_0$~=~(300$\pm$40)~MeV/$c$ 
$\langle |j_{Ty}| \rangle$~=~(190$\pm$25)~MeV/$c$ which is similar although somewhat lower than ISR~\cite{Angelis:1980bs} and RHIC~\cite{Rak:2004gk} measurements. The back-to-back peak broadens with centrality and escapes detection in central collisions. The last measured point ($N_{coll}=$~600) corresponds to $\langle |k_{Ty}| \rangle$~=~(2.8$\pm$0.6)~GeV/$c$ which agrees well with preliminary results from central Au+Au collisions at RHIC~\cite{Rak:2004gk}. 
Within the statistical errors, the yield of the close-angle and back-to-back pion pairs, defined as an area under the Gaussian peak (Fig.~\ref{sigma-yield}b), grows linearly with $N_{coll}$ which supports the suggested interpretation of semihard scattering.

\begin{figure}[t!]
\begin{tabular}{lr}
\includegraphics[height=6.0cm]{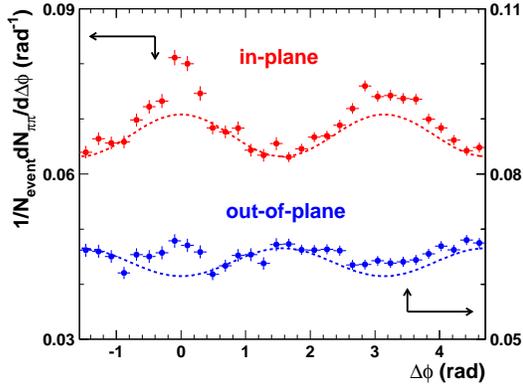}
&
\hspace{-2.5cm}
\begin{minipage}[t]{8.0cm}
\vspace{-5.5cm}
\caption{In-plane (top) and out-of-plane (bottom) two-pion azimuthal correlations for  $p_T>$~1.2~GeV/$c$ and $\Delta\theta>$~20~mrad. Dashed lines are the expectations for pure elliptic flow $v_2$ measured by the event plane method. Data are averaged over centrality classes C1, C2, and C3.}
\label{inout}
\end{minipage}
\end{tabular}
\end{figure}
Due to asymmetry of the overlap zone in non-central collisions, the pion yield might be suppressed if pions propagate perpendicular to the reaction plane rather than along it. We have constructed azimuthal distributions confining one of the pions in the region of $\pm\pi/4$ around the reconstructed event plane ({\it in-plane}) or perpendicular to it ({\it out-of-plane}). The distributions corrected for efficiency losses are shown in Fig.~\ref{inout}. For both in-plane and out-of-plane regions, the data lie above the expectations from elliptic flow~\cite{Bielcikova:2003ku}. After subtracting the flow contributions, we have extracted the ratios of yields in-plane with respect to those out-of-plane. For the close-angle peak this ratio is 1.32$\pm$0.37 and 1.39$\pm$0.44 for the back-to-back component. An additional systematic error of 15$\%$ was estimated due to uncertainties in the subtraction of the elliptic flow contribution. Within the given errors, we can conclude that both components bear only a weak, if any, preference to the reaction plane orientation.

\section{Conclusion}
In summary, we have discussed properties of semihard azimuthal correlations of high-$p_T$ pions embedded 
in collective flow at SPS energy. The observed non-flow components, presumably of semihard origin, show 
similar but  also important differences to observations at RHIC~\cite{Adler:2002tq,Adams:2004wz}. 
The close-angle peak remains narrow at all measured centralities, consistent with fragmentation, 
while the back-to-back component broadens and disappears in background in the most central collisions. 
In addition, there seems to be only a weak, if any, preference of semihard pion pairs to the 
orientation of the reaction plane. This is different from recent findings
at RHIC~\cite{Adams:2004wz} showing a stronger suppression of high-$p_T$ correlations out of the reaction plane.


\section*{References}

\begin{thebibliography}{Bjorken}

\bibitem[1]{Agakichiev:2003gg}
Agakichiev G {\it et al.} (CERES) 2004 \PRL {\bf 92} 032301

\bibitem[2] {Slivova:thesis}
Slivova J 2003 PhD thesis, Charles University, Prague

%\bibitem[3]{Agakishiev:1998wu}
%Agakishiev G {\it et al.} (CERES) 1998 \NP {\bf A638} 467

\bibitem[3]{Lenkeit:1999xu} 
Lenkeit B {\it et al.} (CERES) 1999 \NP {\bf A661} 23


\bibitem[4]{urqmd}
Bass SA {\it et al.} 1998 {\it Prog. Part. Nucl. Phys.} {\bf 41} 225

\bibitem[5]{Eskola:1989yh}
Eskola KJ, Kajantie K and Lindfors J 1989 \NP {\bf B323} 37; 
Miskowiec D, {\tt http://www.gsi.de/\~{}misko/overlap}

\bibitem[6]{Poskanzer:1998} Poskanzer AM, Voloshin SA  1998 \PR {\bf C68} 1671

\bibitem[7]{Dinh:1999mn}
Dinh PM, Borghini N and Ollitrault J-Y 2000  \PL {\bf B477} 51; private communication


\bibitem[8]{Adamova:2002wi}
Adamova D {\it et al.} (CERES) 2003 \NP {\bf A714} 124


\bibitem[9] {RHIC-combined}
Ackermann KH {\it et al.} (STAR) 2001 \PRL {\bf 86} 402; Adcox K {\it et al.} (PHENIX) 2002 \PRL {\bf 89} 212301

\bibitem[10]{Huovinen}
Kolb PF, Huovinen P, Heinz UW, Heiselberg H 2001 \PL {\bf B500} 232; Huovinen P, private communication
%\bibitem[7]{NA49}
%Alt C  {\it et al.} (NA49) 2003 \PR {\bf C68} 034903

\bibitem[11]{Teaney} 
Teaney D nucl-th/0204023; nucl-th/0301099

\bibitem[12]{Rak:2004gk}
Rak J {\it et al.} (PHENIX) 2004 {\it J. Phys.} {\bf G30} S1309-S1312

\bibitem[13]{Angelis:1980bs}
Angelis ALS {\it et al.} (CCOR) 1980 \PL {\bf B97} 163



\bibitem[14]{Bielcikova:2003ku}
Bielcikova J, Esumi S, Filimonov K, Voloshin S and Wurm JP 2004 \PR {\bf C69} 021901 

\bibitem[15]{Adler:2002tq}
Adler C {\it et al.} (STAR) 2003 \PRL {\bf 90} 082302

\bibitem[16]{Adams:2004wz}
Adams J {\it et al.} (STAR) nucl-ex/0407007



\end{thebibliography}
\end{document}